\documentstyle[12pt]{article}
\begin{document}
\renewcommand{\baselinestretch}{1.5}

\newcommand\beq{\begin{equation}}
\newcommand\eeq{\end{equation}}
\newcommand\bea{\begin{eqnarray}}
\newcommand\eea{\end{eqnarray}}

\newcommand\bp{{\bf p}}
\newcommand\bv{{\bf v}}
\newcommand\bA{{\bf A}}
\newcommand\ba{{\bf a}}
\newcommand\br{{\bf r}}
\newcommand\bP{{\bf P}}
\newcommand\bR{{\bf R}}

\newcommand\al{\alpha}
\newcommand\alpi{{\alpha/\pi}}

\newcommand\expo{e^{-{r^2/4l^2}}}
\newcommand\expon{exp(-{r^2/ 4l^2})}
\newcommand\p{\partial}
\newcommand\sumi{\sum_i^N}
\newcommand\sumij{\sum_{i,j \ne i}^N}
\newcommand\sumijk{\sum_{i,j\ne i,k\ne i}^N}
\newcommand\pij{\prod_{i<j}^N}
\newcommand\prl{Phys. Rev. Lett.}
\newcommand\prb{Phys. Rev. {\bf B}}

\hfill MRI-PHY/P971234
\vskip 0.5cm

\centerline{\bf Exact Analytic Results for Composite Fermions}
\centerline{\bf in a Rajaraman-Sondhi like formulation}
\vskip 1 true cm

\centerline{Sumathi Rao \footnote{{\it e-mail
address}: sumathi@mri.ernet.in}} 
\centerline{\it Mehta Research Institute, Chhatnag Road,Jhunsi,}
\centerline{\it Allahabad 211 019, India}
\vskip 2 true cm
\noindent {\bf Abstract}
\vskip 1 true cm

We obtain the exact spectrum and the unique ground state 
of two composite fermions (in a Rajaraman -
Sondhi like formulation) in an external magnetic field $B$. 
We show that the energy eigenvalues decrease with
increasing angular momentum, thus making it energetically favourable 
for composite fermions to stay apart. Generalising this result to a
gas of composite fermions, we provide an energetic justification of
the Laughlin and Jain wave-functions.   
\vskip 1 true cm

\noindent PACS numbers:

\newpage

Experimental discoveries of new heterojunctions\cite{NEW} 
is one reason for the 
continued interest in the phenomenon of the fractional quantum Hall
effect (FQHE)\cite{FQHEBOOK}. 
The other important reason is that theoretically, the
FQHE has been unusually fruitful in giving rise to novel ideas and
excitations, such as composite fermions\cite{JAIN}, 
skyrmions\cite{SONDHI}, a new kind of 
non-Fermi liquid at $\nu=1/2$\cite{HLR}, etc.

Composite fermions are now well-established\cite{EXPT} 
as the relevant quasi-particles
in the fractional quantum Hall effect (FQHE) 
system. True to the nature
of quasi-particles, they are weakly interacting, and in fact, the main
features of the FQHE phenomenon can be simply understood in terms of a model
of non-interacting composite fermions - $i.e.$, FQHE is merely the
integer QHE of non-interacting composite fermions.
Composite fermions were originally introduced in the microscopic,
first quantised trial
wave-function approach\cite{JAIN} by Jain. 
He attached Jastrow factors to 
IQHE wave-functions to get FQHE wave-functions. By interpreting the Jastrow
factors as even units of flux quanta attached to the electrons, he showed 
that FQHE of electrons at fractions $n/(2mn+1)$ is equivalent
to IQHE of composite electrons (electrons with $2m$ flux units
attached) at level $n$.
On a different front, Zhang et al\cite{ZHANG} formulated a field theory
of the FQHE in terms of
a Chern-Simons (CS) gauge field, where the electrons were
interpreted as bosons  with odd number of flux quanta. 
Later, the original composite fermion idea of mapping 
a system of strongly 
interacting fermions in a magnetic field,
to  a system of weakly interacting
composite fermions
was itself implemented 
as  a CS field theory\cite{FRADKIN}.
The composite fermions in the CS model included the phase 
due to the flux quanta 
attached to the fermions, but not its amplitude. A more recent approach
by Rajaraman and Sondhi\cite{RS} remedies this defect, but at the 
expense of making the gauge field complex.

In this letter, we study a system of two composite fermions in an
external magnetic field. 
We use a  Rajaraman-Sondhi like formulation to
model the composite fermions - $i.e.$, our  
composite fermions are ordinary
fermions interacting with the complex vector potential introduced 
by Rajaraman and Sondhi in 
Ref.\cite{RS}.  
Thus, the quantum mechanical problem reduces to that of 
two fermions interacting with a complex vector field and an external
magnetic field. We obtain the energy eigenvalues and the wave-functions
and contrast them with the spectrum obtained by Chern-Simons (CS) 
composite fermions
(composite fermions 
modelled by interaction with a Chern-Simons gauge field). The CS 
composite fermions
behave just like usual 
fermions. They have a large angular momentum degeneracy in the 
presence of an external magnetic field and  no unique ground
state. However, for our two composite fermions, this degeneracy breaks.
The energy decreases as a function of the angular momentum, so the
minimum energy solution is obtained for the maximum value of
angular momentum. For the two composite fermion  system, the maximum is
set by the ratio of the size of the system to the magnetic length.
Hence, the maximum of the angular momentum 
increases with decrease in magnetic length or
equivalently increase in the external magnetic field. We argue,
hence, that the wave-function for many composite fermions should
be an eigenstate of maximum possible angular momentum.

The Hamiltonian for two composite fermions in an external
magnetic field is given by
\beq
H = \sum_i^2{(\bp_i-e\bv_i-e\bA_i)^{\dagger}(\bp_i-e\bv_i-e\bA_i)\over 2m}
\label{hami}  
\eeq
where  $\bA_i=B/2(-y_i,x_i)$ are  the vector potentials of the external 
magnetic field $B$ in the $\hat z$ direction and $\bv_1$ and $\bv_2$
are the Rajaraman-Sondhi (RS) complex gauge fields 
seen by each of the composite
fermions due to the presence of the vortex in the other composite fermion.
As explained in Ref.\cite{RS}, the complex gauge field $\bv_i$ is
related to the usual CS gauge field $\ba_i$ as
\beq
\bv_i = \ba_i + i {\hat z} \times \ba_i. 
\eeq
As is well-known from anyon studies (see for example Ref.\cite{PRIMER}), 
for a 
two particle system, the CS gauge field is given by 
\beq
\ba_1 = {\al\over \pi e} {{\hat z} \times (\br_1-\br_2) \over |\br_1-\br_2|^2},
\quad 
\ba_2 = {\al\over \pi e} {{\hat z} \times (\br_2-\br_1) \over |\br_1-\br_2|^2}.
\eeq
Here $\al$ is the statistics parameter, which for composite fermions is an
even integer $\times\pi$.
The complex gauge fields $\bv_1$ and $\bv_2$ are completely defined
in terms of the CS field. The imaginary term in $\bv_i$ is the radial
component in the gauge field and represents a `fat flux' or spread-out
flux centred at the position of 
the fermion. (The pure CS gauge field, in contrast, attaches infinitesimal
flux-tubes to the fermion thereby only changing its phase.) 

Note that this Hamiltonian is not the same as the Hamiltonian
considered in Ref.\cite{RS}. Their field theoretic 
Hamiltonian was given by
\beq
H_{RS}= \int d^2x \Pi ({\bf x}) {(\bp-e\bv-e\bA)^2\over 2m} 
\chi({\bf x})
\label{hamrs}
\eeq
where $\chi({\bf x})$ denoted the composite fermion field and
$\Pi({\bf x})$ was the  canonically conjugate Fermi field.
However, $\Pi({\bf x})
= \chi^{\dagger}({\bf x})e^{(J({\bf x})+J^{\dagger}({\bf x}))} \neq
\chi^{\dagger}({\bf x})$, where $J({\bf x})$ is related to the 
complex vector potential as $\bv=i{\vec\nabla} J/e$. 
Thus, although $\bv$ is complex, the
Hamiltonian in Eq.(\ref{hamrs}) {\it is} hermitean. 
Note also that the 
RS gauge field in Ref.\cite{RS} also included a $c$-number 
term involving the Gaussian
factor $\expon$, which was needed in their field theoretic formulation,
since the field theory is defined for fixed area. Here, we drop the 
Gaussian factor in the gauge field  
and follow the more common practice of incorporating the whole Gaussian
factor in the IQHE wave-function, with the Gaussian evaluated 
at the external magnetic field\cite{KAMILLA}. Nevertheless, the motivation
for the Hamiltonian in Eq.(\ref{hami}) does come from the RS field theory
and we incorporate its main feature, which is that the gauge field
includes the amplitude as well as the phase of the composite fermions.

The CM motion which just represents a particle with mass $2m$ 
in twice the external magnetic field can be trivially factored out.
We are then left with
the one-particle relative Hamiltonian given by 
\beq
H_{\rm rel} = {(\bp-e\bv-e\bA_{\rm rel})^{\dagger}
(\bp-e\bv-e\bA_{\rm rel})\over m}
\eeq
with 
$\bp = \bp_1-\bp_2$, 
$\bv\equiv \bv_{\rm rel} = \ba_{\rm rel}+i{\hat z}\times \ba_{\rm rel}$
where  
$\ba_{\rm rel}$ in turn is given by $\ba_{\rm rel} = (\al/\pi e) ({\hat z}
\times \br /|\br|^2)$ and $\bA_{\rm rel} = B/4(-y,x)$. 
The wave-function is separable in cylindrical 
coordinates $\psi(r,\theta) = R(r) Y(\theta)$ and we find that
the radial equation reduces to 
\beq
{\large[}-({\p^2\over \p r^2} + {1\over r}{\p\over \p r}) + {1\over r^2} [
(L-\alpi)^2 + (\alpi)^2{\Large]} - {1\over l^2} (L-\alpi) + {r^2\over 4l^4}
-mE{\large]} R(r) = 0 
\eeq
where $l^2 = 2/eB$ is the magnetic length and $L$ is the angular momentum
- $i.e.$,  the angular part of the wave-function is 
$Y(\theta) = e^{iL\theta}$. Fermi statistics of the 
composite fermions require that $L$ be an odd integer. 
The solution of the above radial
equation can be explicitly found in terms of a confluent hypergeometric
function F,
\beq
R(r)=r^s\expo F({1\over 2}
[s - (L-\alpi)+1]-k, 
s+1, {r^2\over 2l^2})
\eeq
with $s=\sqrt{(L-\alpi)^2+(\alpi)^2}$, and $k=mEl^2/2$.
The requirement that the series solution for the confluent hypergeometric 
function terminate leads to the energy eigenvalues
\beq
E_{n,L}= {eB\over 2m}(\sqrt{(L-\alpi)^2+(\alpi)^2}-(L-\alpi)+1+2n)
\label{energy}
\eeq
where $n$ is an integer. For the ground state, $n=0$. 

Before analysing this solution, let us obtain the equivalent solution
when the RS gauge field is replaced by a CS gauge field, for contrast.
The CS Hamiltonian is given by
\beq
H_{\rm rel}^{\rm CS} = H_{\rm rel} - {\al^2\over \pi^2 r^2}
\eeq
The subtracted term on the RHS is  the one that appeared due
to the radial terms in the RS gauge potential. Clearly, the net effect
of the radial term has just been to increase the centrifugal barrier.
For this Hamiltonian,
the solution is simply
\bea
\psi^{\rm CS}(r,\theta)&=& R^{\rm CS}(r)Y^{\rm CS}(\theta) \\
&=& r^{|t|}\expo F({1\over 2}
[|t| - (t)+1]-k, 
|t|+1, {r^2\over 2l^2})  e^{iL\theta}
\eea 
with $t=(L-\alpi)$ yielding the energy eigenvalues 
\beq
E_{n,L}^{\rm CS} = {eB\over 2m}(|L-\alpi|-(L-\alpi)+1+n) 
\eeq
Here, however, by defining a new angular momentum $L'=L-\alpi$, we see
that both the solution (except for an unobservable phase factor, since
for composite fermions, $\alpi$ = even integer) and
the energy eigenvalues reduce to that of an ordinary fermion in an 
external magnetic field. In particular, the massive degeneracy of a
fermion in an external magnetic field 
(all positive values of $L'$ are degenerate with $L'=0$ \footnote{Note
that a change in sign of the external magnetic field changes
the sign of $L'=(L-\alpi)$. For the opposite sign of $B$, all
negative values of $L'$ are degenerate with $L'=0$.})
persists for the CS composite fermion
and there is no unique ground state. This is not surprising, since the 
CS composite fermion is just a gauge transform of the original fermion,
albeit singular - $i.e.$, $\psi_{\rm CS}(\br_1-\br_2)=
(\br_1-\br_2)\psi(\br_1-\br_2)/|(\br_1-\br_2)|$.
(The same result would also be obtained  
if we study a naive first quantised picture
of the RS  field theory. The Hamiltonian is 
$H=\sum_i^2(\bp_i-e\bv_i-e\bA_i)^2/ 2m$ with the hermiticity of the
Hamiltonian being maintained by defining a new inner product in the
Hilbert space 
$
<\psi|O|\phi> = \int \psi^* O \phi 
e^{-(J({\bf x})+J^{\dagger}({\bf x}))} d^2 x,
$
analogous to the field redefinitions made in Ref.\cite{RS}.  The 
eigenvalues of this Hamiltonian also reduce to that of non-interacting
fermions in an external magnetic field and the massive degeneracy 
remains unbroken. The reason, again, is that this Hamiltonian 
can be obtained from the non-interacting Hamiltonian by 
making a transformation - $\psi_{\rm RS}(\br_1-\br_2)
=(\br_1-\br_2)\psi(\br_1-\br_2)$,
although, the transformation is not pure gauge.)

However, the composite fermion interacting with the RS gauge
field as in Eq.(\ref{hami}) is {\it not} merely a  transform of
the non-interacting fermion. This is reflected in the two body problem
explicitly since the wave-functions and energy eigenvalues are now
different. Even after a redefinition of the angular momentum, the radial
part of the wave-function is still different. 
Most interestingly, the massive degeneracy with respect
to angular momentum has disappeared. From the energy expression in 
Eq.(\ref{energy}),
we see that the energy is minimised when $(L-\alpi)$, or equivalently
$L$, is maximised. In the $L\rightarrow\infty$ limit, 
the energy eigenvalue
attains its minimum of $eB/2m$.

For the two CF system that we have solved explicitly, the maximum value
of the angular momentum is fixed by the size of the system . $L_{\rm max}
= mvR$ where $R$ is the size of the system and plugging in the limiting
value of $v$, we obtain $L_{\rm max}=R/l$ - $i.e.$, the size of the
system measured in units of magnetic length. So explicitly for two composite
fermions, we obtain the following result for the ground state -
\bea
R(r)&=&r^s \expo F({1\over 2}
[s' - (L_{\rm max}
-\alpi)+1-mEl^2], 
s'+1, {r^2\over 2l^2})\\
E_0 &=& {eB\over 2m}(s'-
(L_{\rm max}-\alpi)+1)
\eea
where  $s'=\sqrt{(L_{\rm max}-\alpi)^2+(\alpi)^2}$ and $L_{\rm max}=R/l$.

Generalising this result to a system of many composite fermions, we
see that the energy will be minimised if the relative angular momentum
between any pair of composite fermions takes the maximum value that it
can, given the size or equivalently, the density of particles in the
sample. For instance, for FQHE at the fraction $\nu$, the ratio $R/l=1/\nu$,
where we interpret $R$ as the average distance between the composite fermions.
This clearly shows that $L_{\rm max}^{\rm rel}=1/\nu$ and energetically,
this will be the favoured relative angular momentum. A further  assumption of
analyticity, (lowest Landau level condition), leads directly to the 
Laughlin wave-functions. The same argument of maximising relative
angular momentum also justifies the addition of even number of
vortices in the Jain wave-functions. 
(Note however, that the solution for two 
composite fermions is {\it not} analytic because of the square root factor.
Hence, naive generalisation of the wave-function for two composite
fermions does not lead to the correct many-body wave-functions.)

The question of why Coulomb interactions disguise themselves as vortices
attached to the fermions still remains open. However, we have now proved
that if vortices are attached to fermions, then they like to maximise their 
relative angular momenta and stay as far apart as possible, thus 
minimising their Coulomb energy. In other words, we have shown that 
formation of composite fermions mimics the effect of a Coulomb potential
in that it makes it energetically favourable for the fermions to stay
apart. 

We have taken the RS gauge potential as being
the most relevant piece of the RS field theory and studied  
composite fermions in terms of the normal Hamiltonian that one
would write down in the presence of such a complex gauge potential.
Our results, thus, confirm the idea that 
the RS formulation of composite fermions is a much better 
starting point for FQHE than the CS formulation, since the RS gauge field
incorporates Coulomb repulsions. Incidentally, perhaps, 
that is also why the RS formulation is able to obtain the
Laughlin wave-function at the mean field level. Perturbation theory
about the RS mean field FQHE state is  more likely to be 
stable, since incorporation of repulsions implies that the FQHE gap
is already formed at the mean field level. The drawback
is that perturbation theory is more difficult in the RS theory,
due to the non-hermiticity of the mean field Hamiltonian.
Recently, however,  a consistent perturbation theory has been 
formulated\cite{WU}.

In conclusion, let us reiterate the main results of this letter.
We have shown that for two composite fermions in an external magnetic 
field, the maximum possible value of the 
relative angular momentum 
is energetically favoured. This result required the input of the 
amplitude of the vortex attached to the fermion. The pure phase part 
of the vortex, which is what is captured in the CS formulation of
FQHE is not sufficient to break the angular momentum degeneracy.
With this result, it is easy to see why composite fermions
work so well at minimising the Coulomb energy. 
Although, numerically, it
is well-known that composite fermions minimise Coulomb energy, 
this is the first analytic proof of the result.

\section*{Acknowledgments}
We would like to thank J. K. Jain and R. Rajaraman for useful correspondence.


\begin{thebibliography}{99}

\bibitem{NEW} H. L. Stormer et al, \prl ~{\bf 56}, 85 (1986); G. S. Boebinger
et al, \prl ~{\bf 64}, 1793 (1990); Y. W. Suen et al, \prl ~{\bf 68}, 
1379 (1992); J. P. Eisenstein
et al, \prl ~{\bf 68}, 1383 (1992); S. Q. Murphy et al, \prl
~{\bf 72}, 728 (1994).


\bibitem{FQHEBOOK}
D. C. Tsui, H. L. Stormer and A. C. Gossard, \prl ~{\bf 48}, 1559 (1982);
{\it The Quantum Hall Effect}, edited by R. E. Prange and S. M. Girvin,
Springer -Verlag, New York, 1987; {\it The Fractional Quantum Hall Effect},
Springer series in Solid State Sciences, {\bf 85}, Springer-Verlag, 1988.


\bibitem{JAIN} J. K. Jain, \prl ~{\bf 63}, 199 (1989); \prb {\bf 41}, 7653
(1990); Advances in Physics, {\bf 41}, 105 (1992); 
Science {\bf 266}, 1199 (1994).  

\bibitem{SONDHI}
S. L. Sondhi, A. Karkhede, S. A. Kivelson and E. H. Rezayi, \prb
{\bf 47}, 16419 (1993); H. A. Fertig, L. Brey, R. Cote and A. H. 
Macdonald, \prb {\bf 50}, 11018 (1994).


\bibitem{HLR}
B. I. Halperin, P. A. Lee and N. Read, \prb {\bf 47}, 7313 (1993).

\bibitem{EXPT}
W. Kang et al, \prl ~{\bf 71}, 3850 (1993); V. J. Goldman, B. Su amd J. K.
Jain, \prl ~{\bf 72}, 2065 (1994); R. R. Du et al, \prl ~{\bf 75}, 3926.

\bibitem{ZHANG} S. C. Zhang, T. H. Hannson, and S. Kivelson, \prl ~{\bf 62},
82 (1989); S. C. Zhang, Int. J. Mod. Phys. {\bf B 6}, 25 (1992).

\bibitem{FRADKIN} A. Lopez and E. Fradkin, \prb {\bf 44}, 5246 (1991); see
also E. Fradkin, {\it Field Theories of Condensed Matter Systems},
Addison-Wesley, Redwood City, 1991.

\bibitem{RS} R. Rajaraman and S. L. Sondhi, Int. J. of Mod.
Phys. {\bf B 10}, 793 (1996); R. Rajaraman, \prb {\bf 56}, 6788
(1997).

\bibitem{PRIMER}
S. Rao,  `An Anyon Primer', hep-th/9209066, 
published in ``Models and Techniques of Statistical Physics'',
edited by S. M. Bhattacharjee (Narosa Publications). 


\bibitem{KAMILLA}
J. K. Jain and R. K. Kamilla, `Composite fermions : Particles of the 
lowest Landau level', to appear in ``Composite Fermions'' edited by 
Olle Heinonen.

\bibitem{WU}
Y. S. Wu and  Y. Yu, cond-mat/9608061 (1996).

\end{thebibliography}
\end{document}